\documentclass[prl,superscriptaddress,showkeys,showpacs,10pt,tightenlines,floatfix,twocolumn]{revtex4}
\usepackage[utf8]{inputenc}
\usepackage{amsmath}
\usepackage{amsfonts}
\usepackage{amssymb}
\usepackage{graphicx}
\usepackage[normalem]{ulem}

\usepackage[usenames,dvipsnames]{xcolor}
\definecolor{royalazure}{rgb}{0.0, 0.22, 0.66}
\definecolor{rossocorsa}{rgb}{0.83, 0.0, 0.0}
\definecolor{oceanboatblue}{rgb}{0.0, 0.47, 0.75}
\definecolor{azulnormal}{rgb}{0.0, 0.0, 1.00}

\usepackage{hyperref}
\hypersetup{
    colorlinks	= true,
    linkcolor	= rossocorsa,
    filecolor	= magenta,      
    urlcolor	= oceanboatblue,
    citecolor 	= rossocorsa,
}

\usepackage[caption=false]{subfig}

\newcommand{\<}{\left\langle}
\renewcommand{\>}{\right\rangle}

\graphicspath{{./images/}}

\begin{document}

\title{Temporal instability of the frontier between mechanically regulated tissues}

\author{Luis G\'omez-Nava}
\affiliation{Université Paris Cité, Mati\`{e}re et Syst\`{e}mes Complexes (MSC), 10 rue Alice Domon et L\'{e}onie Duquet, 75205 Paris Cedex 13, France.}
\altaffiliation{UMR 7057 CNRS \& Universit\'{e} Paris Cité}
\author{Djamel Chekroun}
\affiliation{Université Paris Cité, Mati\`{e}re et Syst\`{e}mes Complexes (MSC), 10 rue Alice Domon et L\'{e}onie Duquet, 75205 Paris Cedex 13, France.}
\altaffiliation{UMR 7057 CNRS \& Universit\'{e} Paris Cité}
\author{Ioan Ionescu}
\affiliation{Université Sorbonne Paris Nord, Laboratoire des Sciences des Procédés et des Matériaux (LSPM), 99 avenue Jean-Baptiste Clément, 93430 Villetaneuse, France.}
\author{Marc Durand}
\affiliation{Université Paris Cité, Mati\`{e}re et Syst\`{e}mes Complexes (MSC), 10 rue Alice Domon et L\'{e}onie Duquet, 75205 Paris Cedex 13, France.}
\altaffiliation{UMR 7057 CNRS \& Universit\'{e} Paris Cité}
\email{marc.durand@univ-paris-diderot.fr} 
\date{\today}

\newcommand{\revrm}[1]{\textcolor{red}{\sout{#1}}}
\newcommand{\revadd}[1]{\textcolor{blue}{#1}}
\newcommand{\LG}[1]{\textcolor{YellowOrange}{#1}}
\newcommand{\MD}[1]{\textcolor{ForestGreen}{#1}}

\begin{abstract}
	The stability of the boundary between regenerating tissues is essential to the maintenance of their integrity and  biological function. Stress is known to play an important role in the regulation of cell division, cell growth and cell death, and it is thought that stress balance ensures the stability of tissue boundaries. Using a multicellular numerical model, we investigate the stability of the frontier between two confluent cell monolayers whose cell renewal is mechanically regulated. We show that even for two tissues having similar mechanical and biological properties, the location of their common  frontier is subject to strong fluctuations 
	until the complete disappearance of one of the tissues.
	Using a population dynamics model, we show that this temporal instability is inherent to the stochasticity of cell division and cell death events, and derive an analytical expression for the mean disappearance time of a tissue. 
	  These results call for a rethinking of the regulating mechanism of tissue renewal.
\end{abstract}
\maketitle
There is a great interest in understanding how the frontier between a tissue and its environment  \cite{hong2016a,ravasio2015a,basan2013a}, or between two tissues \cite{baker2011a,ninov2007a,vincent2013a,gogna2015a,Podewitz2016,Moitrier_2019}, propagates or maintains a stable shape under cell renewal and activity. 
%
%
%
Propagation of tissue boundaries is observed during morphogenetic events \cite{ninov2010a,bischoff2012a}, wound healing \cite{basan2013a,zimmermann2014a,tarle2015a} or tumour growth \cite{tracqui2009a,cristini2005a,poplawski2010a,ciarletta2011a,amar2011a}. Stabilisation, on the other hand, is mandatory to maintain the integrity and normal functioning of mature tissues.
%
%
There is abundant experimental evidence that cell division and death are mechanically regulated \cite{loza2017a,montel2011a,streichan2014a,puliafito2012a,benham-pyle2015a,aegerter-wilmsen2007a,pan2016a,levayer2016a,gudipaty2017a,fletcher2018a}. This has led naturally to the concept of homeostatic pressure, the pressure at which the rate of cell division is exactly balanced by the rate of cell death \cite{basan_homeostatic_2009}. When the homeostatic pressure of a tissue is high, it means that a high pressure must be applied on it to stop its proliferation.
Various theoretical studies \cite{Williamson2018, Podewitz2016,buscher2020instability,Trenado2021,Li_2022} investigated the competition between  two confluent tissues that differ either in their homeostatic pressure or in their cell motility. Special attention has been carried on the drift and the spatial instability (\emph{fingering}) of the frontier between the two tissues.
%
Within these deterministic models, the position and shape of the frontier are stable when there is no difference in mechanical and biological properties of the two tissues. 
%
%
However, it is well known that stochasticity plays a major role in population dynamics \cite{Lande2003}.
 %
%
 There are two main sources of variability in cell renewal, categorized as \textit{demographic} and \textit{environmental} stochasticity \cite{Engen1998}.
 Demographic stochasticity describes the within-individual variability, while environmental stochasticity refers to temporary environmental fluctuations that lead to changes in population growth rates. These sources of stochasticity are mathematically represented by centered Gaussian noises whose variances depend linearly and quadratically on the actual population size, respectively.
Few studies have examined how the stochasticity of cell renewal or cell motility affects the stability and roughness of tissue boundaries.
Reference \cite{risler_homeostatic_2015} quantifies the fluctuations of the  interface between a tissue and an incompressible medium caused by  the demographic stochasticity of cell renewal. The mean position is assumed to be stable over time.
Reference \cite{williamson_stability_nodate} studies the  propagation and roughness of the frontier between two tissues in presence of noise associated with both cell renewal  and motility. They show that the mean position of the frontier has a standard diffusive behavior, and derive the expression for the diffusive coefficient of the interface position. However, in this approach the variance of the noise was assumed to be independant of the population size.
To our knowledge, the impact of demographic or environmental stochasticity on the boundary stability has never been examined.

In the present Letter, we address this issue and investigate numerically and theoretically how the stochasticity of mechanically regulated cell renewal affects the temporal stability of the frontier between two confluent tissues. For simplicity, we assume that cells have no motility; the fluctuations of the frontier position are caused by cell division (and growth) and cell disappearance only. Furthermore, we consider the idealized situation in which the two cell types have same mechanical 
and biological
properties.
The segregation of the two cell populations is maintained by the extra-cost of adhesive energy between heterotypic contacts compared with homotypic contacts \cite{GranerPRL1992, DurandPLOSCompBiol2021}.
This high boundary energy also inhibits fingering instability \cite{buscher2020instability, Casademunt_fingering}.
If the frontier is unstable in this simple situation, it must also be unstable when the two tissues present some asymmetry in their mechanical or biological properties, for which a drift superimposes on the random fluctuations of the frontier.

Using a standard numerical model for multicellular systems -- the Cellular Potts Model (CPM) -- we show that the frontier between a tissue and an incompressible fluid, or between two tissues, have very different fates: in the former case, the frontier position slightly fluctuates around a stable value, whereas in the latter case the frontier position fluctuates over distances comparable with the system size, until eventually one of the two tissues irremediably disappears.
To have a better understanding of the mechanism that drives this time instability, we then propose a stochastic model for the evolution of the two cell populations, and derive an analytical expression for the mean time over which one of the tissue disappears.
%
%
These results reveal that the widely accepted mechanical regulation mechanism for cell renewal is indeed not sufficient to guarantee the stability and integrity of tissues.
%

%


%
%
%
%
%
%

\paragraph{Numerical model.--}
We numerically investigate the stability of the frontier using the Cellular Potts Model (CPM),  a widely accepted numerical model of multi-cellular systems.
Initially developed to describe single-cell behavior during cell sorting driven by differential adhesion~\cite{GranerPRL1992, GlazierPRE1993}, it was later extended to describe the spatio-temporal evolution of many other collective cellular systems~\cite{VedulaPNAS2012, HirashimaDG&D2017}.
The CPM is a lattice based model: each cell consists of a subset of lattice sites sharing the same cell ID $\sigma$. 
The update algorithm of the CPM is based on a Metropolis-like dynamics~\cite{MombachPRL1995}, introducing an effective temperature which characterizes the amplitude of cell contour fluctuations. We use a version of the algorithm that prevents cell fragmentation \cite{DurandCOMPPHYSCOMM2016}.
%
%
The Hamiltonian of the cellular system reads \cite{GranerPRL1992, GlazierPRE1993}:
\begin{equation}
    \mathcal{H} = \sum_{\substack{neigh. \\ cells \, \langle i, j \rangle}} \gamma_{ij}L_{ij} + \dfrac{B}{2 A_0} \sum_{cells\,\,i} (A_i - A_0)^2.
    \label{eqn:hamiltonian}
\end{equation}
%
%
The first sum in (\ref{eqn:hamiltonian}), computed over neighboring cells $\langle i, j\rangle$, represents the boundary energy.
The coefficient $\gamma_{ij}$ is the energy per unit contact length between neighbouring cells, and depends on the cell types. Here $\gamma_{ij}$ can take two different values, $\gamma_\mathrm{id}$ and $\gamma_\mathrm{diff}$, whether cells $i$, $j$ are of the same type or not. We choose $\gamma_\mathrm{id} < \gamma_\mathrm{diff}$ to ensure that the two tissues do not mix or develop fingering instability. 
The second term in (\ref{eqn:hamiltonian}) represents the compressive energy of the cells, where $B$ is the effective 2D compressive modulus of a cell (which captures its out-of-plane elongation~\cite{VillemotSOFTMATTER2020}), $A_i$ is the area of cell~$i$, and $A_0$ the nominal area. The overpressure of a cell $i$ with nominal (uncompressed) area $A_0$ is then given by $\Pi_i = -\partial \mathcal{H}/\partial A_i = B(1 - A_i/A_0)$. Values of the different parameters are detailed in the S.I..

We incorporated an implementation of mechanically-regulated cell renewal into this numerical model: we impose cells to divide with a constant rate (with daughter cells rapidly growing to the mother cell size). The division rate $\alpha$ is defined as the average fraction of cells in the tissue that divide per unit time (counted in Monte Carlo Sweeps [MCS]). Hence, on regular time intervals $\Delta t$ MCS, every cell divides with probability $\alpha \Delta t$.
%
%
%
Cell disappearance, on the other hand, adjusts as a function of the pressure in the system: when the mean pressure in the tissue exceeds a threshold value $\Pi_c$, cells are randomly removed until the pressure falls below $\Pi_c$ again.
Once a cell is divided (resp. condemned), it quickly grows (resp. shrinks) to the mean (resp. zero) area. Alternatively, we could choose to have a constant apoptosis rate and adjust the cell division and growth depending on the density of cells.
Further technical details on the implementation of the cell division and apoptosis in the CPM are reported in the S.I.
%

\begin{figure}[t]
	\centering
	\includegraphics[width=\columnwidth]{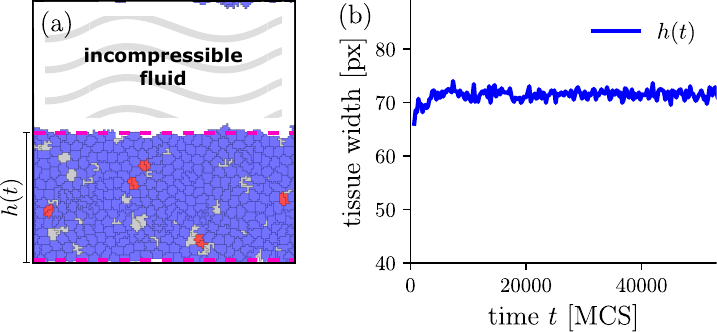}
	\caption{Time evolution of the width of a regenerating tissue in contact with an incompressible fluid. (a) Snapshot of the dynamics at $t=20000$ MCS, where dividing cells are colored in red, and dying cells in light-gray. The frontiers between the tissue and the fluid are highlighted with pink-dashed lines. (b) Temporal evolution of the averaged thickness of the tissue $h(t)$.}
	\label{fig:oneTissue}
\end{figure}

\paragraph{One tissue and an incompressible fluid.--} We first consider the case of a single tissue in contact with an incompressible medium, in a situation comparable to \cite{risler_homeostatic_2015}. In the CPM, the incompressible medium is generated using a single cell with large target area $A_0$ and bulk modulus $B$.
Initially, the tissue and the fluid share equally the available space. 
Due to the periodic boundary conditions employed in the simulations, there are two frontiers between the tissue and the medium.
Figure \ref{fig:oneTissue}a shows the temporal evolution of the spatially averaged width of the tissue $h(t)$, defined as the mean distance between the two frontiers.
After a quick transient regime, $h(t)$ gently fluctuates around a steady state, known as \emph{homeostatic regime}, where the number of disappearing cells equilibrates with the number of dividing cells, in agreement with \cite{risler_homeostatic_2015}. 
%
%

%

\begin{figure}[htb]
	\centering
	\includegraphics[width=\columnwidth]{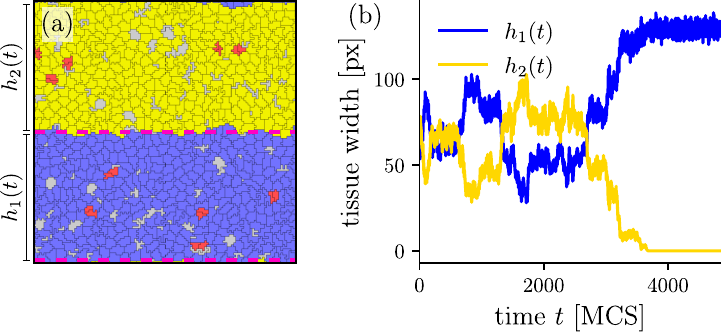}
	\caption{Time evolution of a system of two tissues using the CPM. (a) Snapshot of the dynamics at $t=2000$ MCS, where dividing cells are colored in red and dying cells are colored in light-gray. Cells of type 1 are colored in blue, and cells of type 2 are colored in yellow. The frontiers between the tissues are highlighted with pink-dashed lines. (b) Plot of the thickness of each tissue over time. The numerical simulations were done using a system of $128 \times 128$ sites, a nominal area $A_0=25$, a division rate of $\alpha=0.012$, an apoptosis rate of $\beta=1$ and $r_c=1$\%.}
	\label{fig:TimeEvolution}
\end{figure}

\paragraph{Two tissues.--} We now consider the frontier between two tissues having the same mechanical and biological properties, i.e. equal bulk modulus $B$, target area $A_0$, energy per unit contact length $\gamma_{id}$, division rate $\alpha$ and threshold pressure $\Pi_c$.
Interestingly,, our numerical simulations show that in this case the frontier between the tissues is not stable in time: as shown in Fig. \ref{fig:TimeEvolution}, its mean position fluctuates over distances comparable to the system size, until eventually one or the other tissue disappears.
%
%
%
%

\paragraph{Population Dynamics Model.--}
To gain further understanding of this instability and predict the typical time $T_d$ over which one of the two tissue disappears, we develop a stochastic Population Dynamics Model (PDM), where the temporal evolution of the system is given in terms of the number of cells in the tissues, hence neglecting any advecting contribution.
Assuming that mechanical equilibration within the tissue is fast compared with the division rate, the pressure within the tissue is uniform and the pressure expresses simply in terms of the number of cells in the system $N(t)$:  $\overline  \Pi -\Pi_c=-\frac{B}{A_0}(\overline A  - A_c) \propto 1-N_c/N(t)$, with $N_cA_c=N(t)\overline A $, where the overline indicates averaging over all cells.
In accordance with our CPM simulations, we assume that the division of cells occurs at a constant rate $\alpha$, while cell disappearance is regulated by the overpressure exerted on it: as soon as the number of cells $N(t)$ exceeds $N_c$, $N(t)-N_c$ cells are randomly removed. More conveniently, the dimensionless parameter $r_c=(A_0-A_c)/A_0$ can be used to characterize the critical state. Note that in studies based on a continuous description of tissues \cite{basan_homeostatic_2009,ranft_tissue_2012,williamson_stability_nodate,risler_homeostatic_2015} the pressure is related to the areal density of cells $\rho=1/\overline A$ instead.
%
Neglecting for now the probabilistic nature of cell division and cell disappearance, the evolution of the number of cells in a tissue in contact with an incompressible medium is then determined by the following differential equation:
\begin{equation}
    \dfrac{dN(t)}{dt}=\alpha N(t)-\beta \big( N(t)-N_c \big) \Theta \big(N(t)-N_c\big),
    \label{eqn:pop_dyn_one_tissue_without_noise}
\end{equation}
where $\Theta(x)$ is the Heaviside step function, $\alpha$ is the division rate, and $\beta$ is the rate at which supernumerary cells disappear.
In practice $\beta=1$MCS$^{-1}$ but we keep it in the equation for dimensional purposes.
Suppose $N(0)>N_c$ (if not, there is a short transitory regime during which $N(t)$ grows exponentially with growth rate $\alpha$ until it reaches $N_c$).
From Eq.~(\ref{eqn:pop_dyn_one_tissue_without_noise}), one gets
\begin{equation}
    N(t) = \left(N_0 - N_s \right) e^{-(\beta-\alpha)t} + N_s,
    \label{eqn:solution_one_tissue_without_noise}
\end{equation}
where $N_s=\beta N_c/(\beta - \alpha)$ represents the stationary limit of $N(t)$, which is reached whenever $\alpha < \beta$. 
In the rest of this Letter, we assume that this condition is always satisfied.
The homeostatic pressure
 is then $\Pi_h=\Pi_c+B(A_c/A_0)(\alpha/\beta)$.
We now include the effects of noise associated with the stochasticity of cell division and disappearance events.
Following previous works \cite{Ranft2010,ranft_tissue_2012,risler_homeostatic_2015} we consider here demographic stochasticity, \emph{i.e} stochasticity that occurs because the division and disappearance of each individual is a discrete and probabilistic event (the case of environmental stochasticity will be discussed hereafter). Mathematically, this stochasticity is represented by centered Gaussian white noises with respective variances $\sigma_\alpha^2=\alpha N(t)$ and $\sigma_\beta^2=\beta(N(t)-N_c)$. We suppose here that noises on division and disappearance are uncorrelated, thereby generalizing previous studies in which a single noise is introduced to represent stochasticity of the net division rate (that is, division rate $-$ disappearance rate).
Assuming that $N_0>N_c$, the evolution of the number of cells is given by the following stochastic differential equation (in Ito's formulation):
 \begin{equation}
	\dfrac{dN(t)}{dt}=\alpha N(t)-\beta \big( N(t)-N_c \big)+\sigma_d \xi_d(t)+\sigma_a \xi_\beta(t),
	\label{eqn:pop_dyn_one_tissue_with_noise}
\end{equation}
where $\langle \xi_i(t) \rangle=0$ and  $\langle \xi_i(t) \xi_j(t')\rangle=\delta_{ij}\delta(t-t')$. 
We omitted the Heaviside function as we disregard the transitory regime and suppose that $N(t)>N_c$.
%
One easily shows (see Supplementary Note~S2) that the  average number of cells $\langle N (t)\rangle$ follows the deterministic solution (Eq. \ref{eqn:solution_one_tissue_without_noise}).
Moreover, fluctuations of $N(t)$ around the stationary value $\langle N(t) \rangle = N_s = \beta N_c/(\beta - \alpha)$, characterized by its variance, remain bounded as $(\Delta N)^2={\alpha N_s/(\beta-\alpha)}$.
We corroborated this result by comparing numerical simulations done with the CPM and the solution of the PDM with and without noise, shown in FIG.~S2 and FIG.~S3.
In agreement with \cite{risler_homeostatic_2015}, we conclude that the frontier between a tissue and an incompressible medium is stable: indeed, further growth of cell population increases the pressure above the homeostatic value and favors apoptosis, whereas recession of cell population decreases the pressure and favors division.

\begin{figure}[htbp]
	\centering
	\includegraphics[width=\columnwidth]{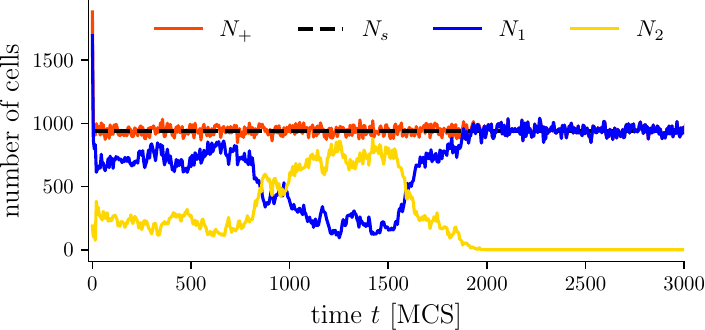}
	\caption{Time evolution of the two cell population and their sum, for a single run.}
	\label{fig:One_run}
\end{figure}

We now extend our model to the case of two cell populations.
We note $N_1(t)$ and $N_2(t)$ the respective number of cells in the two tissues, and $N_+(t)=N_1(t)+N_2(t)$ the total number of cells.
Since the two tissues have the same mechanical and biological properties, the pressure is the same in the two tissues 
and the number of cells that disappear in each tissue are in proportion of their respective populations. 
Hence, equation (\ref{eqn:pop_dyn_one_tissue_with_noise}) generalizes in this case to:
\begin{align}
	\dfrac{dN_1}{dt} & = \alpha N_1(t)-\beta N_1(t)\left(1-{N_c}/{N_+(t)}\right) \nonumber \\ & +\sigma_{1d} \xi_{1d}(t)+\sigma_{1a} \xi_{1a}(t),
    \label{2tissues-noise-a}\\
	\dfrac{dN_2}{dt} & = \alpha N_2(t)-\beta N_2(t)\left(1-{N_c}/{N_+(t)}\right) \nonumber \\ & +\sigma_{2d} \xi_{2d}(t)+\sigma_{2a} \xi_{2a}(t),
	\label{2tissues-noise-b}
\end{align}
where $\sigma_{id}= \sqrt{\alpha N_i}$, $\sigma_{ia}= \sqrt{\beta N_i(1-N_c/N_+)}$ and $\langle \xi_{im}(t) \xi_{jn}(t')\rangle=\delta_{ij}\delta_{mn}\delta(t-t')$.
%
%
The solution to the noise-free version of Eqs (\ref{2tissues-noise-a})-(\ref{2tissues-noise-b}) is $N_1(t)=N_{10}N_+(t)/N_+^0$, $N_2(t)=N_{20}N_+(t)/N_+^0$, with
%
\begin{align}
	N_+(t) = \big( N_+^0 - N_s \big)e^{-(\beta - \alpha)t} + N_s.
	\quad
	\label{2tissues-no_noise-f}
\end{align}
%
%
As happened for a single tissue in contact with an incompressible fluid, the system reaches a stable regime whenever $\beta > \alpha$.
In contrast, the numerical resolution of Eqs (\ref{2tissues-noise-a})-(\ref{2tissues-noise-b}) shows that in presence of noise (see Fig. \ref{fig:One_run}), the cell populations show two different dynamics:  at short times ($t \ll 1/(\beta-\alpha)$)  the dynamics is controlled by the deterministic terms of the equations, the populations follows the same dynamics of their noise-free counterparts until they reach their noise-free steady state values $N_{1s}$, $N_{2s}$. At longer times ($t \gtrsim 1/(\beta-\alpha)$),  the noise terms are not negligible anymore, leading to large fluctuations of the cell populations until one of the two goes extinct.

\begin{figure}[t]
	\centering
	\includegraphics[width=\columnwidth]{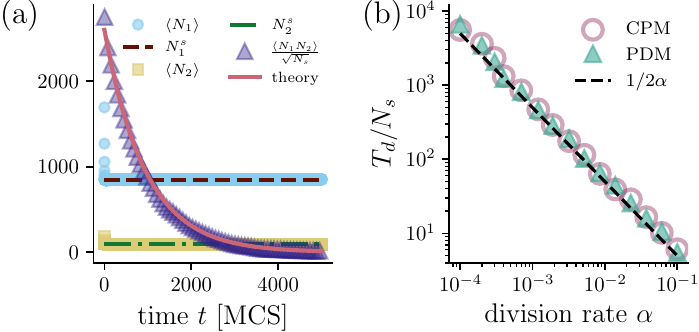}
	\caption{(a) Time evolution of $\<N_1\>$, $\<N_2\>$,  and $\<N_1N_2\>/\sqrt{N_s}$. Comparison between numerical resolution and analytic expression (solid line).
    (b) Normalized average disappearance time $T_d/N_s$ as a function of $\alpha$. Both plots were done using an apoptosis rate of $\beta = 1$, a threshold value $r_c=1$\% and a nominal area $A_0=25$.}
	\label{fig:NMin_Td}
\end{figure}

This simple model thus captures the instability observed in our CPM simulations. Note that while the two populations fluctuate wildly, their sum $N_+$ fluctuates very slightly around $N_s$, its noise-free steady state value.
By performing a large number of runs (with the same initial conditions), we can plot the time evolution of the mean populations. Figure \ref{fig:NMin_Td}a shows that $\< N_1 \>$ and $\< N_2 \>$ reach quickly their steady state values, which seem identical to their noise-free counterparts $N_{is}=N_i^0 N_s/N_+^0$, $i=1,2$.
In contrast, $\< N_1N_2 \>$ is dramatically affected by the presence of noise, as it exponentially decreases  down to $0$ on a timescale that is much longer than $(\beta-\alpha)^{-1}$. This evolution reflects that on each run one or the other population vanishes.
%
%
We can express analytically the long time evolution of $\< N_1N_2(t) \>$: assuming that $N_+(t)=N_s(1+\varepsilon)$ with $\vert \varepsilon \vert \ll 1$), and using Ito's derivation rule, we show (see S.I.) that the evolutions of $\<  N_1N_2 \>$ and $\<  N_1N_2 \varepsilon \>$ are coupled through: 
\begin{align}
	\dfrac{d\< N_1N_2 \>}{dt} & =  \dfrac{2\alpha}{N_s}\< N_1N_2 \>-2(\beta-\alpha)\< N_1N_2 \varepsilon \>,\\
	\dfrac{d\< N_1N_2 \varepsilon\>}{dt} & =  \dfrac{2\alpha}{N_s}\< N_1N_2 \>-(\beta-\alpha)\left(1-\dfrac{2}{N_s}\right)\< N_1N_2 \varepsilon \>.
\end{align}
In the limit $N_s \gg 1$, the two eigenvalues of this linear system are negative, with respective values $\lambda_m=-2\alpha/N_s$ and $\lambda_M=-(\beta-\alpha)+2(\alpha+\beta)/N_s$, with $\vert \lambda_m\vert\ll \vert \lambda_M \vert$ in the thermodynamic limit.
Therefore, the typical time over which $\< N_1N_2\>$ decreases exponentially to 0 is $1/\vert \lambda_m \vert$.

We check the validity of the PDM and its analytic predictions as compared with CPM simulations. Figure~\ref{fig:NMin_Td}b compares the normalized mean disappearance time $T_d/N_s$ as a function of the division rate $\alpha$ both for the CPM and PDM. Both curves 
%
accurately match the predicted theoretical value $1/2\alpha$. It is worth noticing that as $N_s=\beta N_c/(\beta-\alpha)$, the mean disappearance time $T_d$ is actually not linear in $\alpha^{-1}$, and reaches a minimum for $\alpha=\beta/2$. Most importantly, $T_d$ scales linearly with the system size $N_s$, and not with $N_s^2$ as one would expect for a frontier performing a  standard random walk. Because of the mechanically-driven cell disappearance, the effective compressibility of tissues is infinite, and the frontier is subject to giant fluctuations.
%
We also checked that the frontier is still unstable when noise is suppressed either on cell division or disappearance (not shown). Modelling stochasticity of cell division and disappearance by one single noise \cite{Ranft2010,ranft_tissue_2012,risler_homeostatic_2015} can then be viewed as one particular case of the present analysis.
We also checked (see S.I.) that the probability for a tissue to disappear depends on the initial populations, since the probability that the cell population $i$ disappears is given by $1-N_i^0/N_+^0$.

Note that an alternative approach to prove the temporal instability, which is detailed in the S.I., consists in deriving the Fokker-Planck equation corresponding to the Langevin equations (\ref{2tissues-noise-a}) and (\ref{2tissues-noise-b}), and then show that the steady-state distribution is zero whenever $N_1=0$ or $N_2=0$.

\paragraph*{Discussion}
We have shown that the frontier between a tissue and an incompressible fluid, and that between two tissues have very different fates when stochasticity is added to the cell birth/death processes. The former is stable in time with amplitude of fluctuations that goes to zero in the thermodynamic limit. The latter is unstable and shows large fluctuations until one of the two tissues disappears. The mean disappearance time scales linearly with the system size, unlike a diffusive frontier. We may wonder whether this result remains valid if we consider environmental stochasticity instead of demographic stochasticity. Environmental stochasticity occurs \emph{e.g.} when the division/disappearance rates evolve with time. We show numerically and analytically in the  
S.I. that the frontier between two tissues is still unstable, and $\< N_1 N_2\>$ still decreases exponentially, with a rate which depends on $\sigma_d$, $\sigma_a$ and $\alpha$.
%

Our results highlight that mechanically-regulated cell renewal cannot guarantee the stabilization of two competing tissues, and call for a rethinking of the mechanism for stabilizing under renewal. A stabilization mechanism requires the larger population to have a lower renewal rate. Such feedback could occur through the production (or inhibition) of some signaling from dividing/dying cells that is perceived by cells of the same type only.
%
%
As a concluding remark, we stress out that this stability issue is not exclusive to tissues and may occur for other competing biological systems as well (e.g.: bacteria).

\section*{Acknowledgments}
\begin{acknowledgments}
We thank A. Altieri and F. van Wijland for stimulating discussions. ANR (Agence Nationale de la Recherche) and CGI (Commissariat à l’Investissement d’Avenir) are gratefully acknowledged for their financial support of this work through Labex SEAM (Science and Engineering for Advanced Materials and devices), ANR-10-LABX-0096 and ANR-18-IDEX-0001.

\end{acknowledgments}

\bibliography{biblio}

\begin{thebibliography}{48}
\expandafter\ifx\csname natexlab\endcsname\relax\def\natexlab#1{#1}\fi
\expandafter\ifx\csname bibnamefont\endcsname\relax
  \def\bibnamefont#1{#1}\fi
\expandafter\ifx\csname bibfnamefont\endcsname\relax
  \def\bibfnamefont#1{#1}\fi
\expandafter\ifx\csname citenamefont\endcsname\relax
  \def\citenamefont#1{#1}\fi
\expandafter\ifx\csname url\endcsname\relax
  \def\url#1{\texttt{#1}}\fi
\expandafter\ifx\csname urlprefix\endcsname\relax\def\urlprefix{URL }\fi
\providecommand{\bibinfo}[2]{#2}
\providecommand{\eprint}[2][]{\url{#2}}

\bibitem[{\citenamefont{Hong et~al.}(2016)\citenamefont{Hong, Dumond, Tsugawa,
  Sapala, Routier-Kierzkowska, Zhou, Chen, Kiss, Zhu, Hamant
  et~al.}}]{hong2016a}
\bibinfo{author}{\bibfnamefont{L.}~\bibnamefont{Hong}},
  \bibinfo{author}{\bibfnamefont{M.}~\bibnamefont{Dumond}},
  \bibinfo{author}{\bibfnamefont{S.}~\bibnamefont{Tsugawa}},
  \bibinfo{author}{\bibfnamefont{A.}~\bibnamefont{Sapala}},
  \bibinfo{author}{\bibfnamefont{A.-L.} \bibnamefont{Routier-Kierzkowska}},
  \bibinfo{author}{\bibfnamefont{Y.}~\bibnamefont{Zhou}},
  \bibinfo{author}{\bibfnamefont{C.}~\bibnamefont{Chen}},
  \bibinfo{author}{\bibfnamefont{A.}~\bibnamefont{Kiss}},
  \bibinfo{author}{\bibfnamefont{M.}~\bibnamefont{Zhu}},
  \bibinfo{author}{\bibfnamefont{O.}~\bibnamefont{Hamant}},
  \bibnamefont{et~al.}, \bibinfo{journal}{Dev. Cell}
  \textbf{\bibinfo{volume}{38}}, \bibinfo{pages}{15} (\bibinfo{year}{2016}).

\bibitem[{\citenamefont{Ravasio et~al.}(2015)\citenamefont{Ravasio, Cheddadi,
  Chen, Pereira, Ong, Bertocchi, Brugues, Jacinto, Kabla, Toyama
  et~al.}}]{ravasio2015a}
\bibinfo{author}{\bibfnamefont{A.}~\bibnamefont{Ravasio}},
  \bibinfo{author}{\bibfnamefont{I.}~\bibnamefont{Cheddadi}},
  \bibinfo{author}{\bibfnamefont{T.}~\bibnamefont{Chen}},
  \bibinfo{author}{\bibfnamefont{T.}~\bibnamefont{Pereira}},
  \bibinfo{author}{\bibfnamefont{H.}~\bibnamefont{Ong}},
  \bibinfo{author}{\bibfnamefont{C.}~\bibnamefont{Bertocchi}},
  \bibinfo{author}{\bibfnamefont{A.}~\bibnamefont{Brugues}},
  \bibinfo{author}{\bibfnamefont{A.}~\bibnamefont{Jacinto}},
  \bibinfo{author}{\bibfnamefont{A.}~\bibnamefont{Kabla}},
  \bibinfo{author}{\bibfnamefont{Y.}~\bibnamefont{Toyama}},
  \bibnamefont{et~al.}, \bibinfo{journal}{Nat. Commun}
  \textbf{\bibinfo{volume}{6}}, \bibinfo{pages}{7683} (\bibinfo{year}{2015}).

\bibitem[{\citenamefont{Basan et~al.}(2013)\citenamefont{Basan, Elgeti,
  Hannezo, Rappel, and Levine}}]{basan2013a}
\bibinfo{author}{\bibfnamefont{M.}~\bibnamefont{Basan}},
  \bibinfo{author}{\bibfnamefont{J.}~\bibnamefont{Elgeti}},
  \bibinfo{author}{\bibfnamefont{E.}~\bibnamefont{Hannezo}},
  \bibinfo{author}{\bibfnamefont{W.}~\bibnamefont{Rappel}}, \bibnamefont{and}
  \bibinfo{author}{\bibfnamefont{H.}~\bibnamefont{Levine}},
  \bibinfo{journal}{Proc. Natl. Acad. Sci. U.S. A}
  \textbf{\bibinfo{volume}{110}}, \bibinfo{pages}{2452} (\bibinfo{year}{2013}).

\bibitem[{\citenamefont{Baker}(2011)}]{baker2011a}
\bibinfo{author}{\bibfnamefont{N.}~\bibnamefont{Baker}},
  \bibinfo{journal}{Curr. Biol} \textbf{\bibinfo{volume}{21, R11}}
  (\bibinfo{year}{2011}).

\bibitem[{\citenamefont{Ninov and Martin-Blanco}(2007)}]{ninov2007a}
\bibinfo{author}{\bibfnamefont{N.}~\bibnamefont{Ninov}} \bibnamefont{and}
  \bibinfo{author}{\bibfnamefont{E.}~\bibnamefont{Martin-Blanco}},
  \bibinfo{journal}{Nat. Protoc} \textbf{\bibinfo{volume}{2}},
  \bibinfo{pages}{3074} (\bibinfo{year}{2007}).

\bibitem[{\citenamefont{Vincent et~al.}(2013)\citenamefont{Vincent, Fletcher,
  and Baena-Lopez}}]{vincent2013a}
\bibinfo{author}{\bibfnamefont{J.-P.} \bibnamefont{Vincent}},
  \bibinfo{author}{\bibfnamefont{A.}~\bibnamefont{Fletcher}}, \bibnamefont{and}
  \bibinfo{author}{\bibfnamefont{L.}~\bibnamefont{Baena-Lopez}},
  \bibinfo{journal}{Nat. Rev. Mol. Cell Biol} \textbf{\bibinfo{volume}{14}},
  \bibinfo{pages}{581} (\bibinfo{year}{2013}).

\bibitem[{\citenamefont{Gogna et~al.}(2015)\citenamefont{Gogna, Shee, and
  Moreno}}]{gogna2015a}
\bibinfo{author}{\bibfnamefont{R.}~\bibnamefont{Gogna}},
  \bibinfo{author}{\bibfnamefont{K.}~\bibnamefont{Shee}}, \bibnamefont{and}
  \bibinfo{author}{\bibfnamefont{E.}~\bibnamefont{Moreno}},
  \bibinfo{journal}{Annu. Rev. Genet} \textbf{\bibinfo{volume}{49}},
  \bibinfo{pages}{697} (\bibinfo{year}{2015}).

\bibitem[{\citenamefont{Podewitz et~al.}(2016)\citenamefont{Podewitz,
  Jülicher, Gompper, and Elgeti}}]{Podewitz2016}
\bibinfo{author}{\bibfnamefont{N.}~\bibnamefont{Podewitz}},
  \bibinfo{author}{\bibfnamefont{F.}~\bibnamefont{Jülicher}},
  \bibinfo{author}{\bibfnamefont{G.}~\bibnamefont{Gompper}}, \bibnamefont{and}
  \bibinfo{author}{\bibfnamefont{J.}~\bibnamefont{Elgeti}},
  \bibinfo{journal}{New Journal of Physics} \textbf{\bibinfo{volume}{18}},
  \bibinfo{pages}{083020} (\bibinfo{year}{2016}), ISSN
  \bibinfo{issn}{1367-2630},
  \urlprefix\url{https://iopscience.iop.org/article/10.1088/1367-2630/18/8/083020}.

\bibitem[{\citenamefont{Moitrier et~al.}(2019)\citenamefont{Moitrier,
  Blanch-Mercader, Garcia, Sliogeryte, Martin, Camonis, Marcq, Silberzan, and
  Bonnet}}]{Moitrier_2019}
\bibinfo{author}{\bibfnamefont{S.}~\bibnamefont{Moitrier}},
  \bibinfo{author}{\bibfnamefont{C.}~\bibnamefont{Blanch-Mercader}},
  \bibinfo{author}{\bibfnamefont{S.}~\bibnamefont{Garcia}},
  \bibinfo{author}{\bibfnamefont{K.}~\bibnamefont{Sliogeryte}},
  \bibinfo{author}{\bibfnamefont{T.}~\bibnamefont{Martin}},
  \bibinfo{author}{\bibfnamefont{J.}~\bibnamefont{Camonis}},
  \bibinfo{author}{\bibfnamefont{P.}~\bibnamefont{Marcq}},
  \bibinfo{author}{\bibfnamefont{P.}~\bibnamefont{Silberzan}},
  \bibnamefont{and} \bibinfo{author}{\bibfnamefont{I.}~\bibnamefont{Bonnet}},
  \bibinfo{journal}{Soft Matter} \textbf{\bibinfo{volume}{15}},
  \bibinfo{pages}{537} (\bibinfo{year}{2019}),
  \urlprefix\url{http://dx.doi.org/10.1039/C8SM01523F}.

\bibitem[{\citenamefont{Ninov et~al.}(2010)\citenamefont{Ninov, Menezes-Cabral,
  Prat-Rojo, Manjón, Weiss, Pyrowolakis, Affolter, and
  MartIn-Blanco}}]{ninov2010a}
\bibinfo{author}{\bibfnamefont{N.}~\bibnamefont{Ninov}},
  \bibinfo{author}{\bibfnamefont{S.}~\bibnamefont{Menezes-Cabral}},
  \bibinfo{author}{\bibfnamefont{C.}~\bibnamefont{Prat-Rojo}},
  \bibinfo{author}{\bibfnamefont{C.}~\bibnamefont{Manjón}},
  \bibinfo{author}{\bibfnamefont{A.}~\bibnamefont{Weiss}},
  \bibinfo{author}{\bibfnamefont{G.}~\bibnamefont{Pyrowolakis}},
  \bibinfo{author}{\bibfnamefont{M.}~\bibnamefont{Affolter}}, \bibnamefont{and}
  \bibinfo{author}{\bibfnamefont{E.}~\bibnamefont{MartIn-Blanco}},
  \bibinfo{journal}{Curr. Biol} \textbf{\bibinfo{volume}{20}},
  \bibinfo{pages}{513} (\bibinfo{year}{2010}).

\bibitem[{\citenamefont{Bischoff}(2012)}]{bischoff2012a}
\bibinfo{author}{\bibfnamefont{M.}~\bibnamefont{Bischoff}},
  \bibinfo{journal}{Dev. Biol} \textbf{\bibinfo{volume}{363}},
  \bibinfo{pages}{179} (\bibinfo{year}{2012}).

\bibitem[{\citenamefont{Zimmermann et~al.}(2014)\citenamefont{Zimmermann,
  Basan, and Levine}}]{zimmermann2014a}
\bibinfo{author}{\bibfnamefont{J.}~\bibnamefont{Zimmermann}},
  \bibinfo{author}{\bibfnamefont{M.}~\bibnamefont{Basan}}, \bibnamefont{and}
  \bibinfo{author}{\bibfnamefont{H.}~\bibnamefont{Levine}},
  \bibinfo{journal}{Eur. Phys. J. Spec. Top} \textbf{\bibinfo{volume}{223}},
  \bibinfo{pages}{1259} (\bibinfo{year}{2014}).

\bibitem[{\citenamefont{Tarle et~al.}(2015)\citenamefont{Tarle, Ravasio, Hakim,
  and Gov}}]{tarle2015a}
\bibinfo{author}{\bibfnamefont{V.}~\bibnamefont{Tarle}},
  \bibinfo{author}{\bibfnamefont{A.}~\bibnamefont{Ravasio}},
  \bibinfo{author}{\bibfnamefont{V.}~\bibnamefont{Hakim}}, \bibnamefont{and}
  \bibinfo{author}{\bibfnamefont{N.}~\bibnamefont{Gov}},
  \bibinfo{journal}{Integr. Biol} \textbf{\bibinfo{volume}{7}},
  \bibinfo{pages}{1218} (\bibinfo{year}{2015}).

\bibitem[{\citenamefont{Tracqui}(2009)}]{tracqui2009a}
\bibinfo{author}{\bibfnamefont{P.}~\bibnamefont{Tracqui}},
  \bibinfo{journal}{Rep. Prog. Phys} \textbf{\bibinfo{volume}{72}},
  \bibinfo{pages}{056701} (\bibinfo{year}{2009}).

\bibitem[{\citenamefont{Cristini et~al.}(2005)\citenamefont{Cristini, Frieboes,
  Gatenby, Caserta, Ferrari, and Sinek}}]{cristini2005a}
\bibinfo{author}{\bibfnamefont{V.}~\bibnamefont{Cristini}},
  \bibinfo{author}{\bibfnamefont{H.}~\bibnamefont{Frieboes}},
  \bibinfo{author}{\bibfnamefont{R.}~\bibnamefont{Gatenby}},
  \bibinfo{author}{\bibfnamefont{S.}~\bibnamefont{Caserta}},
  \bibinfo{author}{\bibfnamefont{M.}~\bibnamefont{Ferrari}}, \bibnamefont{and}
  \bibinfo{author}{\bibfnamefont{J.}~\bibnamefont{Sinek}},
  \bibinfo{journal}{Clin. Cancer Res} \textbf{\bibinfo{volume}{11}},
  \bibinfo{pages}{6772} (\bibinfo{year}{2005}).

\bibitem[{\citenamefont{Poplawski et~al.}(2010)\citenamefont{Poplawski,
  Shirinifard, Agero, Gens, Swat, and Glazier}}]{poplawski2010a}
\bibinfo{author}{\bibfnamefont{N.}~\bibnamefont{Poplawski}},
  \bibinfo{author}{\bibfnamefont{A.}~\bibnamefont{Shirinifard}},
  \bibinfo{author}{\bibfnamefont{U.}~\bibnamefont{Agero}},
  \bibinfo{author}{\bibfnamefont{J.}~\bibnamefont{Gens}},
  \bibinfo{author}{\bibfnamefont{M.}~\bibnamefont{Swat}}, \bibnamefont{and}
  \bibinfo{author}{\bibfnamefont{J.}~\bibnamefont{Glazier}},
  \bibinfo{journal}{PLoS One} \textbf{\bibinfo{volume}{5}},
  \bibinfo{pages}{10641} (\bibinfo{year}{2010}).

\bibitem[{\citenamefont{Ciarletta et~al.}(2011)\citenamefont{Ciarletta, Foret,
  and Amar}}]{ciarletta2011a}
\bibinfo{author}{\bibfnamefont{P.}~\bibnamefont{Ciarletta}},
  \bibinfo{author}{\bibfnamefont{L.}~\bibnamefont{Foret}}, \bibnamefont{and}
  \bibinfo{author}{\bibfnamefont{M.}~\bibnamefont{Amar}}, \bibinfo{journal}{J.
  R. Soc. Interface} \textbf{\bibinfo{volume}{8}}, \bibinfo{pages}{345}
  (\bibinfo{year}{2011}).

\bibitem[{\citenamefont{Amar et~al.}(2011)\citenamefont{Amar, Chatelain, and
  Ciarletta}}]{amar2011a}
\bibinfo{author}{\bibfnamefont{M.}~\bibnamefont{Amar}},
  \bibinfo{author}{\bibfnamefont{C.}~\bibnamefont{Chatelain}},
  \bibnamefont{and}
  \bibinfo{author}{\bibfnamefont{P.}~\bibnamefont{Ciarletta}},
  \bibinfo{journal}{Phys. Rev. Lett} \textbf{\bibinfo{volume}{106}},
  \bibinfo{pages}{148101} (\bibinfo{year}{2011}).

\bibitem[{\citenamefont{Loza and Thompson}(2017)}]{loza2017a}
\bibinfo{author}{\bibfnamefont{M.}~\bibnamefont{Loza}} \bibnamefont{and}
  \bibinfo{author}{\bibfnamefont{B.}~\bibnamefont{Thompson}},
  \bibinfo{journal}{Mech. Dev. A} \textbf{\bibinfo{volume}{144}},
  \bibinfo{pages}{23} (\bibinfo{year}{2017}).

\bibitem[{\citenamefont{Montel et~al.}(2011)\citenamefont{Montel, Delarue,
  Elgeti, Malaquin, Basan, Risler, Cabane, Vignjevic, Prost, Cappello
  et~al.}}]{montel2011a}
\bibinfo{author}{\bibfnamefont{F.}~\bibnamefont{Montel}},
  \bibinfo{author}{\bibfnamefont{M.}~\bibnamefont{Delarue}},
  \bibinfo{author}{\bibfnamefont{J.}~\bibnamefont{Elgeti}},
  \bibinfo{author}{\bibfnamefont{L.}~\bibnamefont{Malaquin}},
  \bibinfo{author}{\bibfnamefont{M.}~\bibnamefont{Basan}},
  \bibinfo{author}{\bibfnamefont{T.}~\bibnamefont{Risler}},
  \bibinfo{author}{\bibfnamefont{B.}~\bibnamefont{Cabane}},
  \bibinfo{author}{\bibfnamefont{D.}~\bibnamefont{Vignjevic}},
  \bibinfo{author}{\bibfnamefont{J.}~\bibnamefont{Prost}},
  \bibinfo{author}{\bibfnamefont{G.}~\bibnamefont{Cappello}},
  \bibnamefont{et~al.}, \bibinfo{journal}{Phys. Rev. Lett}
  \textbf{\bibinfo{volume}{107}}, \bibinfo{pages}{188102}
  (\bibinfo{year}{2011}).

\bibitem[{\citenamefont{Streichan et~al.}(2014)\citenamefont{Streichan,
  Hoerner, Schneidt, Holzer, and Hufnagel}}]{streichan2014a}
\bibinfo{author}{\bibfnamefont{S.}~\bibnamefont{Streichan}},
  \bibinfo{author}{\bibfnamefont{C.}~\bibnamefont{Hoerner}},
  \bibinfo{author}{\bibfnamefont{T.}~\bibnamefont{Schneidt}},
  \bibinfo{author}{\bibfnamefont{D.}~\bibnamefont{Holzer}}, \bibnamefont{and}
  \bibinfo{author}{\bibfnamefont{L.}~\bibnamefont{Hufnagel}},
  \bibinfo{journal}{Proc. Natl. Acad. Sci. U.S.A}
  \textbf{\bibinfo{volume}{111}}, \bibinfo{pages}{5586} (\bibinfo{year}{2014}).

\bibitem[{\citenamefont{Puliafito et~al.}(2012)\citenamefont{Puliafito,
  Hufnagel, Neveu, Streichan, Sigal, Fygenson, and Shraiman}}]{puliafito2012a}
\bibinfo{author}{\bibfnamefont{A.}~\bibnamefont{Puliafito}},
  \bibinfo{author}{\bibfnamefont{L.}~\bibnamefont{Hufnagel}},
  \bibinfo{author}{\bibfnamefont{P.}~\bibnamefont{Neveu}},
  \bibinfo{author}{\bibfnamefont{S.}~\bibnamefont{Streichan}},
  \bibinfo{author}{\bibfnamefont{A.}~\bibnamefont{Sigal}},
  \bibinfo{author}{\bibfnamefont{D.}~\bibnamefont{Fygenson}}, \bibnamefont{and}
  \bibinfo{author}{\bibfnamefont{B.}~\bibnamefont{Shraiman}},
  \bibinfo{journal}{Proc. Natl. Acad. Sci. U.S.A}
  \textbf{\bibinfo{volume}{109}}, \bibinfo{pages}{739} (\bibinfo{year}{2012}).

\bibitem[{\citenamefont{Benham-Pyle et~al.}(2015)\citenamefont{Benham-Pyle,
  Pruitt, and Nelson}}]{benham-pyle2015a}
\bibinfo{author}{\bibfnamefont{B.}~\bibnamefont{Benham-Pyle}},
  \bibinfo{author}{\bibfnamefont{B.}~\bibnamefont{Pruitt}}, \bibnamefont{and}
  \bibinfo{author}{\bibfnamefont{W.}~\bibnamefont{Nelson}},
  \bibinfo{journal}{Science} \textbf{\bibinfo{volume}{348}},
  \bibinfo{pages}{1024} (\bibinfo{year}{2015}).

\bibitem[{\citenamefont{Aegerter-Wilmsen
  et~al.}(2007)\citenamefont{Aegerter-Wilmsen, Aegerter, Hafen, and
  Basler}}]{aegerter-wilmsen2007a}
\bibinfo{author}{\bibfnamefont{T.}~\bibnamefont{Aegerter-Wilmsen}},
  \bibinfo{author}{\bibfnamefont{C.}~\bibnamefont{Aegerter}},
  \bibinfo{author}{\bibfnamefont{E.}~\bibnamefont{Hafen}}, \bibnamefont{and}
  \bibinfo{author}{\bibfnamefont{K.}~\bibnamefont{Basler}},
  \bibinfo{journal}{Mech. Dev} \textbf{\bibinfo{volume}{124}},
  \bibinfo{pages}{318} (\bibinfo{year}{2007}).

\bibitem[{\citenamefont{Pan et~al.}(2016)\citenamefont{Pan, Heemskerk, Ibar,
  Shraiman, and Irvine}}]{pan2016a}
\bibinfo{author}{\bibfnamefont{Y.}~\bibnamefont{Pan}},
  \bibinfo{author}{\bibfnamefont{I.}~\bibnamefont{Heemskerk}},
  \bibinfo{author}{\bibfnamefont{C.}~\bibnamefont{Ibar}},
  \bibinfo{author}{\bibfnamefont{B.}~\bibnamefont{Shraiman}}, \bibnamefont{and}
  \bibinfo{author}{\bibfnamefont{K.}~\bibnamefont{Irvine}},
  \bibinfo{journal}{Proc. Natl. Acad. Sci. U.S.A}
  \textbf{\bibinfo{volume}{113}}, \bibinfo{pages}{6974} (\bibinfo{year}{2016}).

\bibitem[{\citenamefont{Levayer et~al.}(2016)\citenamefont{Levayer, Dupont, and
  Moreno}}]{levayer2016a}
\bibinfo{author}{\bibfnamefont{R.}~\bibnamefont{Levayer}},
  \bibinfo{author}{\bibfnamefont{C.}~\bibnamefont{Dupont}}, \bibnamefont{and}
  \bibinfo{author}{\bibfnamefont{E.}~\bibnamefont{Moreno}},
  \bibinfo{journal}{Curr. Biol} \textbf{\bibinfo{volume}{26}},
  \bibinfo{pages}{670} (\bibinfo{year}{2016}).

\bibitem[{\citenamefont{Gudipaty et~al.}(2017)\citenamefont{Gudipaty, Lindblom,
  Loftus, Redd, Edes, Davey, Krishnegowda, and Rosenblatt}}]{gudipaty2017a}
\bibinfo{author}{\bibfnamefont{S.}~\bibnamefont{Gudipaty}},
  \bibinfo{author}{\bibfnamefont{J.}~\bibnamefont{Lindblom}},
  \bibinfo{author}{\bibfnamefont{P.}~\bibnamefont{Loftus}},
  \bibinfo{author}{\bibfnamefont{M.}~\bibnamefont{Redd}},
  \bibinfo{author}{\bibfnamefont{K.}~\bibnamefont{Edes}},
  \bibinfo{author}{\bibfnamefont{C.}~\bibnamefont{Davey}},
  \bibinfo{author}{\bibfnamefont{V.}~\bibnamefont{Krishnegowda}},
  \bibnamefont{and}
  \bibinfo{author}{\bibfnamefont{J.}~\bibnamefont{Rosenblatt}},
  \bibinfo{journal}{Nature} \textbf{\bibinfo{volume}{543}},
  \bibinfo{pages}{118} (\bibinfo{year}{2017}).

\bibitem[{\citenamefont{Fletcher et~al.}(2018)\citenamefont{Fletcher, Loza,
  Borreguero-Muñoz, Holder, Aguilar-Aragon, and Thompson}}]{fletcher2018a}
\bibinfo{author}{\bibfnamefont{G.}~\bibnamefont{Fletcher}},
  \bibinfo{author}{\bibfnamefont{M.-d.-C.-d.-l.} \bibnamefont{Loza}},
  \bibinfo{author}{\bibfnamefont{N.}~\bibnamefont{Borreguero-Muñoz}},
  \bibinfo{author}{\bibfnamefont{M.}~\bibnamefont{Holder}},
  \bibinfo{author}{\bibfnamefont{M.}~\bibnamefont{Aguilar-Aragon}},
  \bibnamefont{and} \bibinfo{author}{\bibfnamefont{B.}~\bibnamefont{Thompson}},
  \bibinfo{journal}{Development} \textbf{\bibinfo{volume}{145}},
  \bibinfo{pages}{159467} (\bibinfo{year}{2018}).

\bibitem[{\citenamefont{Basan et~al.}(2009)\citenamefont{Basan, Risler, Joanny,
  Sastre‐Garau, and Prost}}]{basan_homeostatic_2009}
\bibinfo{author}{\bibfnamefont{M.}~\bibnamefont{Basan}},
  \bibinfo{author}{\bibfnamefont{T.}~\bibnamefont{Risler}},
  \bibinfo{author}{\bibfnamefont{J.}~\bibnamefont{Joanny}},
  \bibinfo{author}{\bibfnamefont{X.}~\bibnamefont{Sastre‐Garau}},
  \bibnamefont{and} \bibinfo{author}{\bibfnamefont{J.}~\bibnamefont{Prost}},
  \bibinfo{journal}{HFSP Journal} \textbf{\bibinfo{volume}{3}},
  \bibinfo{pages}{265} (\bibinfo{year}{2009}), ISSN \bibinfo{issn}{1955-2068},
  \urlprefix\url{https://www.tandfonline.com/doi/full/10.2976/1.3086732}.

\bibitem[{\citenamefont{Williamson and
  Salbreux}(2018{\natexlab{a}})}]{Williamson2018}
\bibinfo{author}{\bibfnamefont{J.~J.} \bibnamefont{Williamson}}
  \bibnamefont{and} \bibinfo{author}{\bibfnamefont{G.}~\bibnamefont{Salbreux}},
  \bibinfo{journal}{Physical Review Letters} \textbf{\bibinfo{volume}{121}},
  \bibinfo{pages}{238102} (\bibinfo{year}{2018}{\natexlab{a}}), ISSN
  \bibinfo{issn}{0031-9007, 1079-7114},
  \urlprefix\url{https://link.aps.org/doi/10.1103/PhysRevLett.121.238102}.

\bibitem[{\citenamefont{B{\"u}scher et~al.}(2020)\citenamefont{B{\"u}scher,
  Diez, Gompper, and Elgeti}}]{buscher2020instability}
\bibinfo{author}{\bibfnamefont{T.}~\bibnamefont{B{\"u}scher}},
  \bibinfo{author}{\bibfnamefont{A.~L.} \bibnamefont{Diez}},
  \bibinfo{author}{\bibfnamefont{G.}~\bibnamefont{Gompper}}, \bibnamefont{and}
  \bibinfo{author}{\bibfnamefont{J.}~\bibnamefont{Elgeti}},
  \bibinfo{journal}{New journal of physics} \textbf{\bibinfo{volume}{22}},
  \bibinfo{pages}{083005} (\bibinfo{year}{2020}).

\bibitem[{\citenamefont{Trenado et~al.}(2021)\citenamefont{Trenado, Bonilla,
  and Martínez-Calvo}}]{Trenado2021}
\bibinfo{author}{\bibfnamefont{C.}~\bibnamefont{Trenado}},
  \bibinfo{author}{\bibfnamefont{L.~L.} \bibnamefont{Bonilla}},
  \bibnamefont{and}
  \bibinfo{author}{\bibfnamefont{A.}~\bibnamefont{Martínez-Calvo}},
  \bibinfo{journal}{Soft Matter} \textbf{\bibinfo{volume}{17}},
  \bibinfo{pages}{8276} (\bibinfo{year}{2021}),
  \urlprefix\url{http://dx.doi.org/10.1039/D1SM00626F}.

\bibitem[{\citenamefont{Li et~al.}(2022)\citenamefont{Li, Schnyder, Turner, and
  Yamamoto}}]{Li_2022}
\bibinfo{author}{\bibfnamefont{J.}~\bibnamefont{Li}},
  \bibinfo{author}{\bibfnamefont{S.~K.} \bibnamefont{Schnyder}},
  \bibinfo{author}{\bibfnamefont{M.~S.} \bibnamefont{Turner}},
  \bibnamefont{and} \bibinfo{author}{\bibfnamefont{R.}~\bibnamefont{Yamamoto}},
  \bibinfo{journal}{Phys. Rev. Res.} \textbf{\bibinfo{volume}{4}},
  \bibinfo{pages}{033156} (\bibinfo{year}{2022}),
  \urlprefix\url{https://link.aps.org/doi/10.1103/PhysRevResearch.4.033156}.

\bibitem[{\citenamefont{Lande et~al.}(2003)\citenamefont{Lande, Engen, and
  Saether}}]{Lande2003}
\bibinfo{author}{\bibfnamefont{R.}~\bibnamefont{Lande}},
  \bibinfo{author}{\bibfnamefont{S.}~\bibnamefont{Engen}}, \bibnamefont{and}
  \bibinfo{author}{\bibfnamefont{B.-E.} \bibnamefont{Saether}},
  \emph{\bibinfo{title}{Stochastic population dynamics in ecology and
  conservation}} (\bibinfo{publisher}{Oxford University Press, USA},
  \bibinfo{year}{2003}).

\bibitem[{\citenamefont{Engen et~al.}(1998)\citenamefont{Engen, Øyvind Bakke,
  and Islam}}]{Engen1998}
\bibinfo{author}{\bibfnamefont{S.}~\bibnamefont{Engen}},
  \bibinfo{author}{\bibnamefont{Øyvind Bakke}}, \bibnamefont{and}
  \bibinfo{author}{\bibfnamefont{A.}~\bibnamefont{Islam}},
  \bibinfo{journal}{Biometrics} \textbf{\bibinfo{volume}{54}},
  \bibinfo{pages}{840} (\bibinfo{year}{1998}), ISSN \bibinfo{issn}{0006341X,
  15410420}, \urlprefix\url{http://www.jstor.org/stable/2533838}.

\bibitem[{\citenamefont{Risler et~al.}(2015)\citenamefont{Risler, Peilloux, and
  Prost}}]{risler_homeostatic_2015}
\bibinfo{author}{\bibfnamefont{T.}~\bibnamefont{Risler}},
  \bibinfo{author}{\bibfnamefont{A.}~\bibnamefont{Peilloux}}, \bibnamefont{and}
  \bibinfo{author}{\bibfnamefont{J.}~\bibnamefont{Prost}},
  \bibinfo{journal}{Physical Review Letters} \textbf{\bibinfo{volume}{115}},
  \bibinfo{pages}{258104} (\bibinfo{year}{2015}), ISSN
  \bibinfo{issn}{0031-9007, 1079-7114},
  \urlprefix\url{https://link.aps.org/doi/10.1103/PhysRevLett.115.258104}.

\bibitem[{\citenamefont{Williamson and
  Salbreux}(2018{\natexlab{b}})}]{williamson_stability_nodate}
\bibinfo{author}{\bibfnamefont{J.~J.} \bibnamefont{Williamson}}
  \bibnamefont{and} \bibinfo{author}{\bibfnamefont{G.}~\bibnamefont{Salbreux}},
  p.~\bibinfo{pages}{17} (\bibinfo{year}{2018}{\natexlab{b}}).

\bibitem[{\citenamefont{Graner and Glazier}(1992)}]{GranerPRL1992}
\bibinfo{author}{\bibfnamefont{F.}~\bibnamefont{Graner}} \bibnamefont{and}
  \bibinfo{author}{\bibfnamefont{J.~A.} \bibnamefont{Glazier}},
  \bibinfo{journal}{Phys. Rev. Lett.} \textbf{\bibinfo{volume}{69}},
  \bibinfo{pages}{2013} (\bibinfo{year}{1992}),
  \urlprefix\url{https://link.aps.org/doi/10.1103/PhysRevLett.69.2013}.

\bibitem[{\citenamefont{Durand}(2021)}]{DurandPLOSCompBiol2021}
\bibinfo{author}{\bibfnamefont{M.}~\bibnamefont{Durand}},
  \bibinfo{journal}{PLOS Computational Biology} \textbf{\bibinfo{volume}{17}},
  \bibinfo{pages}{1} (\bibinfo{year}{2021}),
  \urlprefix\url{https://doi.org/10.1371/journal.pcbi.1008576}.

\bibitem[{\citenamefont{Alert et~al.}(2019)\citenamefont{Alert,
  Blanch-Mercader, and Casademunt}}]{Casademunt_fingering}
\bibinfo{author}{\bibfnamefont{R.}~\bibnamefont{Alert}},
  \bibinfo{author}{\bibfnamefont{C.}~\bibnamefont{Blanch-Mercader}},
  \bibnamefont{and}
  \bibinfo{author}{\bibfnamefont{J.}~\bibnamefont{Casademunt}},
  \bibinfo{journal}{Phys. Rev. Lett.} \textbf{\bibinfo{volume}{122}},
  \bibinfo{pages}{088104} (\bibinfo{year}{2019}),
  \urlprefix\url{https://link.aps.org/doi/10.1103/PhysRevLett.122.088104}.

\bibitem[{\citenamefont{Glazier and Graner}(1993)}]{GlazierPRE1993}
\bibinfo{author}{\bibfnamefont{J.~A.} \bibnamefont{Glazier}} \bibnamefont{and}
  \bibinfo{author}{\bibfnamefont{F.~c.} \bibnamefont{Graner}},
  \bibinfo{journal}{Phys. Rev. E} \textbf{\bibinfo{volume}{47}},
  \bibinfo{pages}{2128} (\bibinfo{year}{1993}),
  \urlprefix\url{https://link.aps.org/doi/10.1103/PhysRevE.47.2128}.

\bibitem[{\citenamefont{Vedula et~al.}(2012)\citenamefont{Vedula, Leong, Lai,
  Hersen, Kabla, Lim, and Ladoux}}]{VedulaPNAS2012}
\bibinfo{author}{\bibfnamefont{S.~R.~K.} \bibnamefont{Vedula}},
  \bibinfo{author}{\bibfnamefont{M.~C.} \bibnamefont{Leong}},
  \bibinfo{author}{\bibfnamefont{T.~L.} \bibnamefont{Lai}},
  \bibinfo{author}{\bibfnamefont{P.}~\bibnamefont{Hersen}},
  \bibinfo{author}{\bibfnamefont{A.~J.} \bibnamefont{Kabla}},
  \bibinfo{author}{\bibfnamefont{C.~T.} \bibnamefont{Lim}}, \bibnamefont{and}
  \bibinfo{author}{\bibfnamefont{B.}~\bibnamefont{Ladoux}},
  \bibinfo{journal}{Proceedings of the National Academy of Sciences}
  \textbf{\bibinfo{volume}{109}}, \bibinfo{pages}{12974}
  (\bibinfo{year}{2012}),
  \eprint{https://www.pnas.org/doi/pdf/10.1073/pnas.1119313109},
  \urlprefix\url{https://www.pnas.org/doi/abs/10.1073/pnas.1119313109}.

\bibitem[{\citenamefont{Hirashima et~al.}(2017)\citenamefont{Hirashima, Rens,
  and Merks}}]{HirashimaDG&D2017}
\bibinfo{author}{\bibfnamefont{T.}~\bibnamefont{Hirashima}},
  \bibinfo{author}{\bibfnamefont{E.~G.} \bibnamefont{Rens}}, \bibnamefont{and}
  \bibinfo{author}{\bibfnamefont{R.~M.~H.} \bibnamefont{Merks}},
  \bibinfo{journal}{Development, Growth \& Differentiation}
  \textbf{\bibinfo{volume}{59}}, \bibinfo{pages}{329} (\bibinfo{year}{2017}),
  \eprint{https://onlinelibrary.wiley.com/doi/pdf/10.1111/dgd.12358},
  \urlprefix\url{https://onlinelibrary.wiley.com/doi/abs/10.1111/dgd.12358}.

\bibitem[{\citenamefont{Mombach et~al.}(1995)\citenamefont{Mombach, Glazier,
  Raphael, and Zajac}}]{MombachPRL1995}
\bibinfo{author}{\bibfnamefont{J.~C.~M.} \bibnamefont{Mombach}},
  \bibinfo{author}{\bibfnamefont{J.~A.} \bibnamefont{Glazier}},
  \bibinfo{author}{\bibfnamefont{R.~C.} \bibnamefont{Raphael}},
  \bibnamefont{and} \bibinfo{author}{\bibfnamefont{M.}~\bibnamefont{Zajac}},
  \bibinfo{journal}{Phys. Rev. Lett.} \textbf{\bibinfo{volume}{75}},
  \bibinfo{pages}{2244} (\bibinfo{year}{1995}),
  \urlprefix\url{https://link.aps.org/doi/10.1103/PhysRevLett.75.2244}.

\bibitem[{\citenamefont{Durand and Guesnet}(2016)}]{DurandCOMPPHYSCOMM2016}
\bibinfo{author}{\bibfnamefont{M.}~\bibnamefont{Durand}} \bibnamefont{and}
  \bibinfo{author}{\bibfnamefont{E.}~\bibnamefont{Guesnet}},
  \bibinfo{journal}{Computer Physics Communications}
  \textbf{\bibinfo{volume}{208}}, \bibinfo{pages}{54} (\bibinfo{year}{2016}),
  ISSN \bibinfo{issn}{0010-4655},
  \urlprefix\url{https://www.sciencedirect.com/science/article/pii/S0010465516302284}.

\bibitem[{\citenamefont{Villemot et~al.}(2020)\citenamefont{Villemot,
  Calmettes, and Durand}}]{VillemotSOFTMATTER2020}
\bibinfo{author}{\bibfnamefont{F.}~\bibnamefont{Villemot}},
  \bibinfo{author}{\bibfnamefont{A.}~\bibnamefont{Calmettes}},
  \bibnamefont{and} \bibinfo{author}{\bibfnamefont{M.}~\bibnamefont{Durand}},
  \bibinfo{journal}{Soft Matter} \textbf{\bibinfo{volume}{16}},
  \bibinfo{pages}{10358} (\bibinfo{year}{2020}),
  \urlprefix\url{http://dx.doi.org/10.1039/D0SM01113D}.

\bibitem[{\citenamefont{Ranft et~al.}(2012)\citenamefont{Ranft, Prost,
  Jülicher, and Joanny}}]{ranft_tissue_2012}
\bibinfo{author}{\bibfnamefont{J.}~\bibnamefont{Ranft}},
  \bibinfo{author}{\bibfnamefont{J.}~\bibnamefont{Prost}},
  \bibinfo{author}{\bibfnamefont{F.}~\bibnamefont{Jülicher}},
  \bibnamefont{and} \bibinfo{author}{\bibfnamefont{J.~F.}
  \bibnamefont{Joanny}}, \bibinfo{journal}{The European Physical Journal E}
  \textbf{\bibinfo{volume}{35}}, \bibinfo{pages}{46} (\bibinfo{year}{2012}),
  ISSN \bibinfo{issn}{1292-8941, 1292-895X},
  \urlprefix\url{http://link.springer.com/10.1140/epje/i2012-12046-5}.

\bibitem[{\citenamefont{Ranft et~al.}(2010)\citenamefont{Ranft, Basan, Elgeti,
  Joanny, Prost, and Jülicher}}]{Ranft2010}
\bibinfo{author}{\bibfnamefont{J.}~\bibnamefont{Ranft}},
  \bibinfo{author}{\bibfnamefont{M.}~\bibnamefont{Basan}},
  \bibinfo{author}{\bibfnamefont{J.}~\bibnamefont{Elgeti}},
  \bibinfo{author}{\bibfnamefont{J.-F.} \bibnamefont{Joanny}},
  \bibinfo{author}{\bibfnamefont{J.}~\bibnamefont{Prost}}, \bibnamefont{and}
  \bibinfo{author}{\bibfnamefont{F.}~\bibnamefont{Jülicher}},
  \bibinfo{journal}{Proceedings of the National Academy of Sciences}
  \textbf{\bibinfo{volume}{107}}, \bibinfo{pages}{20863}
  (\bibinfo{year}{2010}), ISSN \bibinfo{issn}{0027-8424, 1091-6490},
  \urlprefix\url{https://pnas.org/doi/full/10.1073/pnas.1011086107}.

\end{thebibliography}


\end{document}